\documentclass[letterpaper,prl,twocolumn,showpacs,superscriptaddress]{revtex4}

\usepackage{graphicx}
\usepackage{bm}

\newcommand{\eg} {{e.g., }}

\newcommand{\ie} {{i.e., }}
\newcommand{\rmB} {{\rm B}}

\newcommand{\rmd} {{\rm d}}

\newcommand{\rmL} {{\rm L}}
\newcommand{\rms} {{\rm s}}
\newcommand{\rmT} {{\rm T}}
\newcommand{\vecf} {{\bf f}}
\newcommand{\vecr} {{\bf r}}
\newcommand{\vecrho} {\bm{\rho}}

\newcommand{\vecx} {{\bf x}}
\newcommand{\vecy} {{\bf y}}

\begin{document}

\title
{Anomalous hydrodynamic interaction in a quasi-two-dimensional 
suspension}

\author{Bianxiao Cui}
\altaffiliation[Present address: ]
{Department of Physics, Stanford University,
Stanford, California 94305}
\affiliation{Department of Chemistry and The James Franck Institute,
The University of Chicago, Chicago, Illinois 60637}

\author{Haim Diamant}
\email{hdiamant@tau.ac.il}
\affiliation{School of Chemistry, Raymond and Beverly Sackler Faculty
of Exact Sciences, Tel Aviv University,
Tel Aviv 69978, Israel}

\author{Binhua Lin}
\affiliation{The James Franck Institute and CARS,
The University of Chicago, Chicago, Illinois 60637}

\author{Stuart A.\ Rice}
\affiliation{Department of Chemistry and The James Franck Institute,
The University of Chicago, Chicago, Illinois 60637}

\date{March 28, 2004}

\begin{abstract}
  We have studied the correlated Brownian motion of micron-size
  particles suspended in water and confined between two plates. The
  hydrodynamic interaction between the particles exhibits three
  anomalies. (i) The transverse coupling is negative, \ie particles
  exert ``anti-drag'' on one another when moving perpendicular to
  their connecting line. (ii) The interaction decays with
  inter-particle distance $r$ as $1/r^2$, faster than in unconfined
  suspensions but slower than near a single wall. (iii) At large
  distances the pair interaction is independent of concentration
  within the experimental accuracy. The confined suspension thus
  provides an unusual example of long-range, yet essentially pairwise
  correlations even at high concentration.  These effects are shown to
  arise from the two-dimensional dipolar form of the flow induced by
  single-particle motion.
\end{abstract}


\maketitle

Liquid suspensions containing particles of nanometer-to-micron size
(colloids) are ubiquitous in industry and biology \cite{Russel}. In
various circumstances the colloids are spatially confined,
\eg in porous media, biological constrictions, 
nozzles, and microfluidic devices \cite{Whitesides}. The dynamic
behavior of suspensions is affected by flow-mediated velocity
correlations. In unconfined suspensions \cite{Russel} these
hydrodynamic interactions are positive (particles drag one another in
the same direction), long-ranged (decaying with inter-particle
distance $r$ as $1/r$), and involve many-body effects (\ie depend on
concentration)
\cite{Chaikin}.
Considerable efforts have been devoted recently to studying colloids
at the microscopic level, particularly in confinement,
using digital video microscopy \cite{DVM} and optical tweezers
\cite{tweezers}. Particle dynamics near a
single wall \cite{1par1wall,2par1wall}, between two walls
\cite{1par2wall}, in a linear channel
\cite{PRL02}, and 
in a finite container
\cite{Chaikin} were investigated.  These studies have highlighted
flow-mediated effects of the boundaries on particle dynamics.
Quasi-two-dimensional (q-2D) suspensions 
have been studied recently also by computer simulations
\cite{Nagele}.
In this Letter we demonstrate, using digital video microscopy, that
the hydrodynamic interactions in a q-2D suspension are
drastically different from those in less confined as well as more
confined systems.

\begin{figure}[tbh]
\centerline{\resizebox{0.23\textwidth}{!}
{\includegraphics{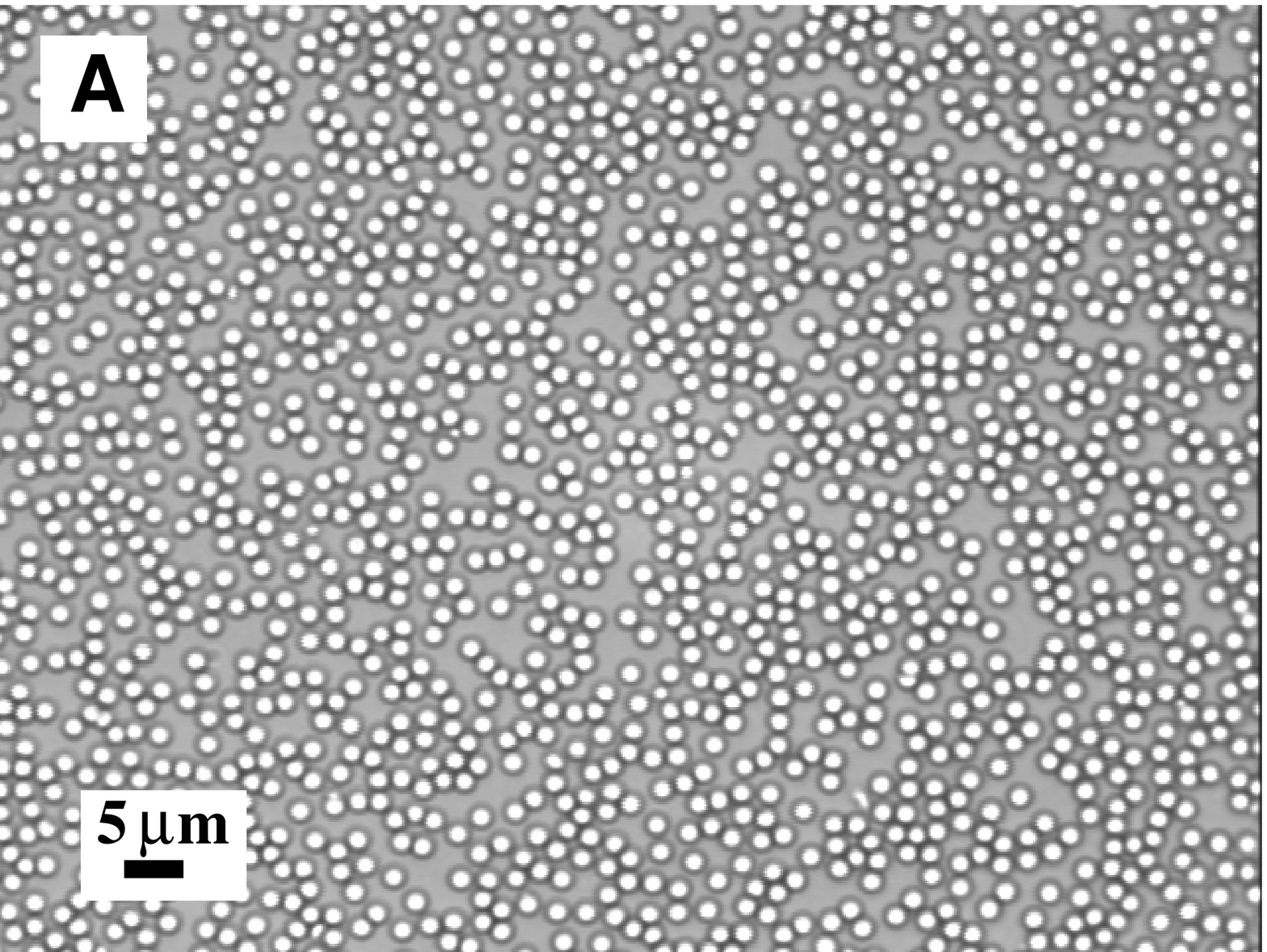}}
\hspace{0cm}
\resizebox{0.23\textwidth}{!}
{\includegraphics{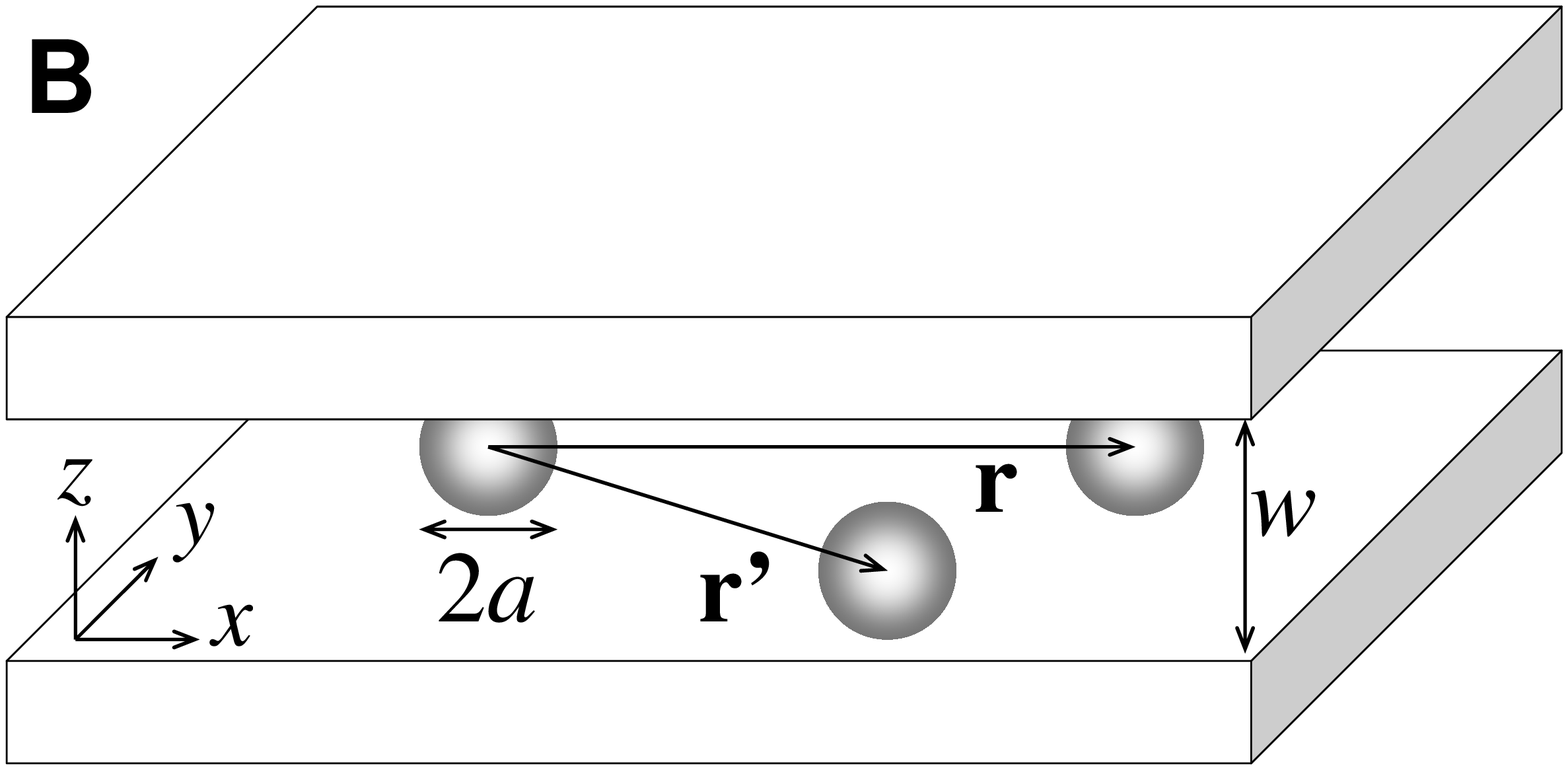}}}
\caption[]{(a) Optical microscope image 
of 1.58 $\mu$m-diameter silica spheres suspended in water and 
confined between two glass plates at area fraction $\phi=0.338$.
(b) Schematic view of the system and its parameters.}
\label{fig_system}
\end{figure}

The experimental system consists of an aqueous suspension of
monodisperse silica spheres (diameter $2a=1.58\pm 0.04$ $\mu$m,
density 2.2 g/cm$^3$, Duke Scientific), tightly confined between two
parallel glass plates in a sealed thin cell (Fig.\ \ref{fig_system}).
The inter-plate separation is $w=1.76\pm 0.05$ $\mu$m \cite{note_w},
\ie slightly larger than the sphere diameter, $2a/w\simeq 0.90$.
Particles thus move essentially parallel to the plates, forming a q-2D
suspension. The glass surfaces are coated with chlorine-terminated
polydimethylsiloxane telomer (Glassclad 6C, United Chemical Products)
to avoid particle sticking.  Digital video microscopy and subsequent
data analysis are used to locate the centers of the $N$ spheres in the
field of view (area $A=106\times 80$ $\mu$m$^2$) and then extract
time-dependent two-dimensional trajectories (time resolution 0.033 s).
Details of the setup and measurement methods have been described
elsewhere \cite{JCP01}.  Measurements were made at four packing
fractions, $\phi\equiv N\pi a^2/A=$ 0.254, 0.338, 0.547, $0.619\pm
0.001$. From equilibrium studies of this system \cite{JCP02} we infer
that, for the the current study, the particles can be
regarded as hard spheres.

For a pair of particles in two dimensions, separated by a distance
$r$, the coupled Brownian motion can be characterized by four
diffusion coefficients, denoted $D_\rmL^\pm(r)$ and
$D_\rmT^\pm(r)$. The longitudinal and transverse collective
coefficients, $D_\rmL^+$ and $D_\rmT^+$, characterize the diffusion
of the center of mass of the pair along and perpendicular to
their connecting line. The relative coefficients, $D_\rmL^-$ and
$D_\rmT^-$, relate to the fluctuations in length and orientation of
the vector $\vecr$ connecting the pair. These coefficients
are measured as
$D_\rmL^\pm(r)=\langle[x_1(t)\pm x_2(t)]^2\rangle_r/(4t)$ and
$D_\rmT^\pm(r)=\langle[y_1(t)\pm y_2(t)]^2\rangle_r/(4t)$,
where $x_i(t)$ and $y_i(t)$ are the displacements of particle $i$
during the time interval $t$ along and perpendicular to the line
connecting it to its partner. The average $\langle\cdots\rangle_r$ is
taken over all observed pairs whose mutual distance falls in the range
$r\pm 0.09$ $\mu$m. The time interval was kept sufficiently short
($t<0.2$ s) that $r$ could be assumed constant within the $\pm
0.09$ $\mu$m bin size. The pair diffusion coefficients are defined 
above such that, in the absence of correlations (\eg when
$r\rightarrow\infty$), all four reduce to the
self-diffusivity of a single particle, $D_\rms$. (This is also
verified experimentally.) Thus, the deviations of $D_{\rmL,\rmT}^\pm$
from $D_\rms$ serve as our measure of the pair hydrodynamic
interaction.

\begin{figure}[tbh]
\centerline{\resizebox{0.4\textwidth}{!}
{\includegraphics{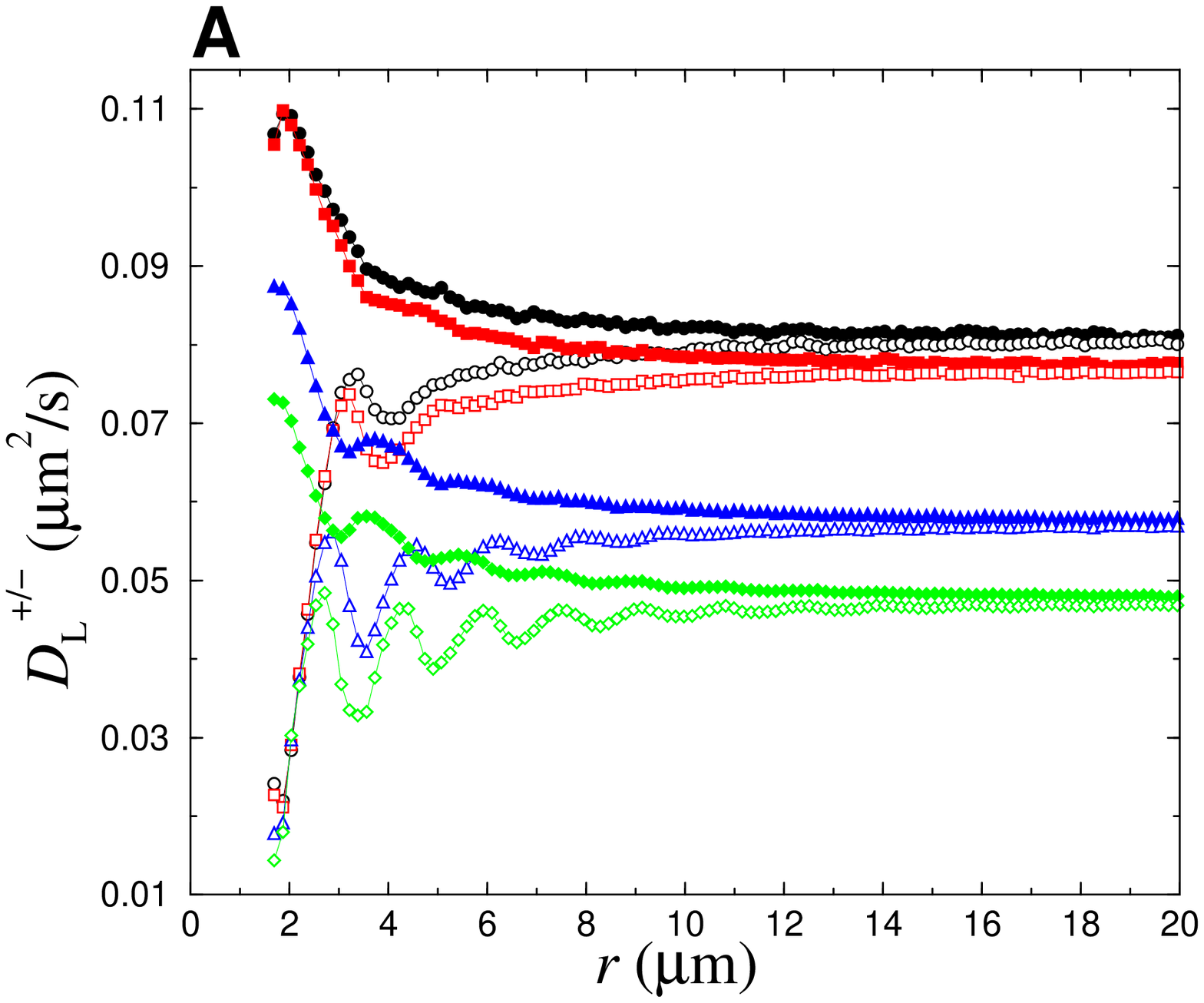}}}
\vspace{-.5cm}
\centerline{
\resizebox{0.4\textwidth}{!}
{\includegraphics{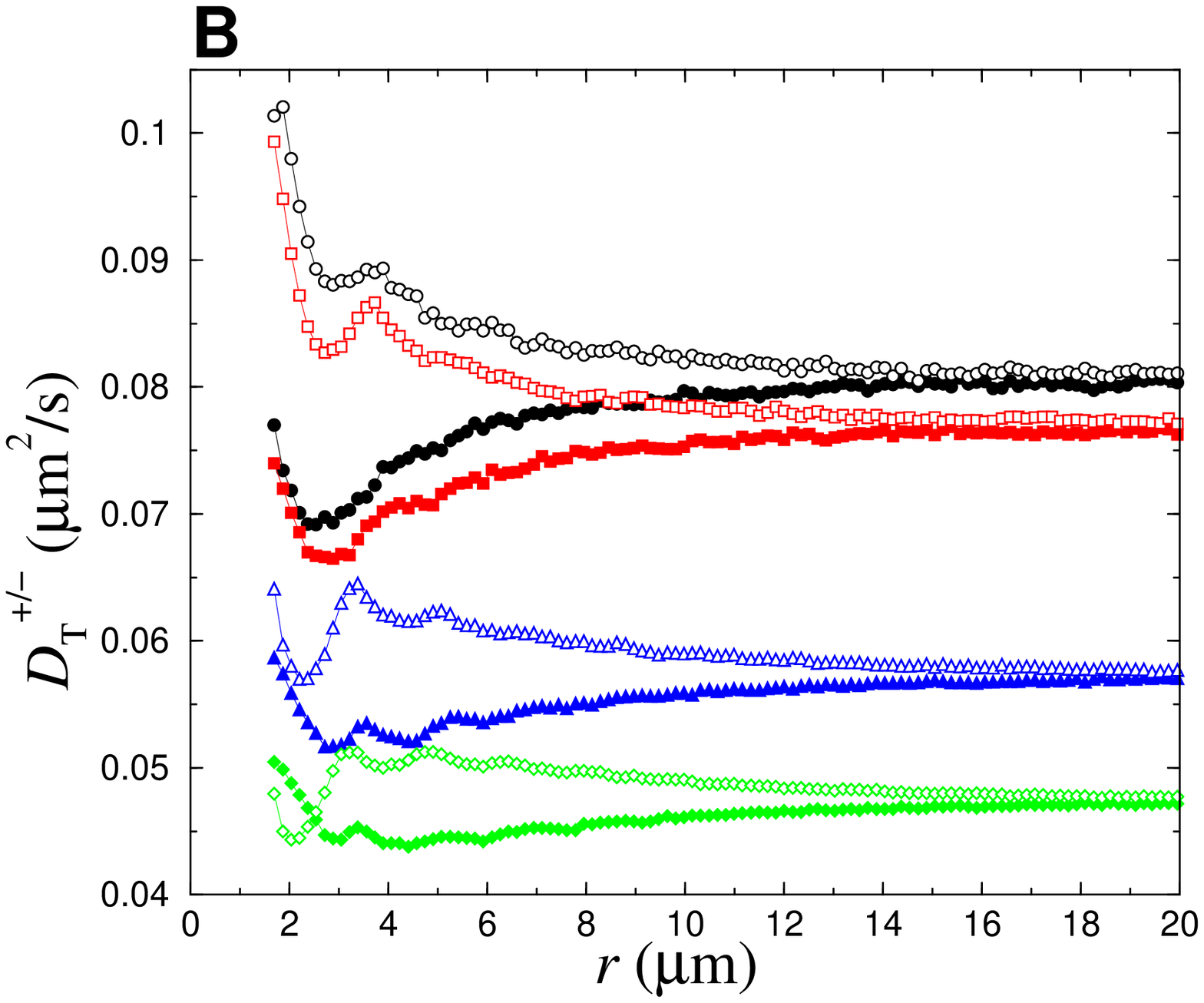}}}
\vspace{-0.5cm}
\caption[]{(Color online)
Longitudinal (a) and transverse (b) pair diffusion coefficients as
functions of inter-particle distance. Filled and open symbols
correspond, respectively, to collective and relative coefficients.
Symbol shapes and colors correspond to different area fractions:
$\phi=$ 0.254 (black circles), 0.338 (red squares), 0.547 (blue
triangles), 0.619 (green diamonds). Note the long range of the
interaction and the sign reversal of the transverse coupling.}
\label{fig_D}
\end{figure}

\begin{figure}[tbh]
\centerline{\resizebox{0.4\textwidth}{!}
{\includegraphics{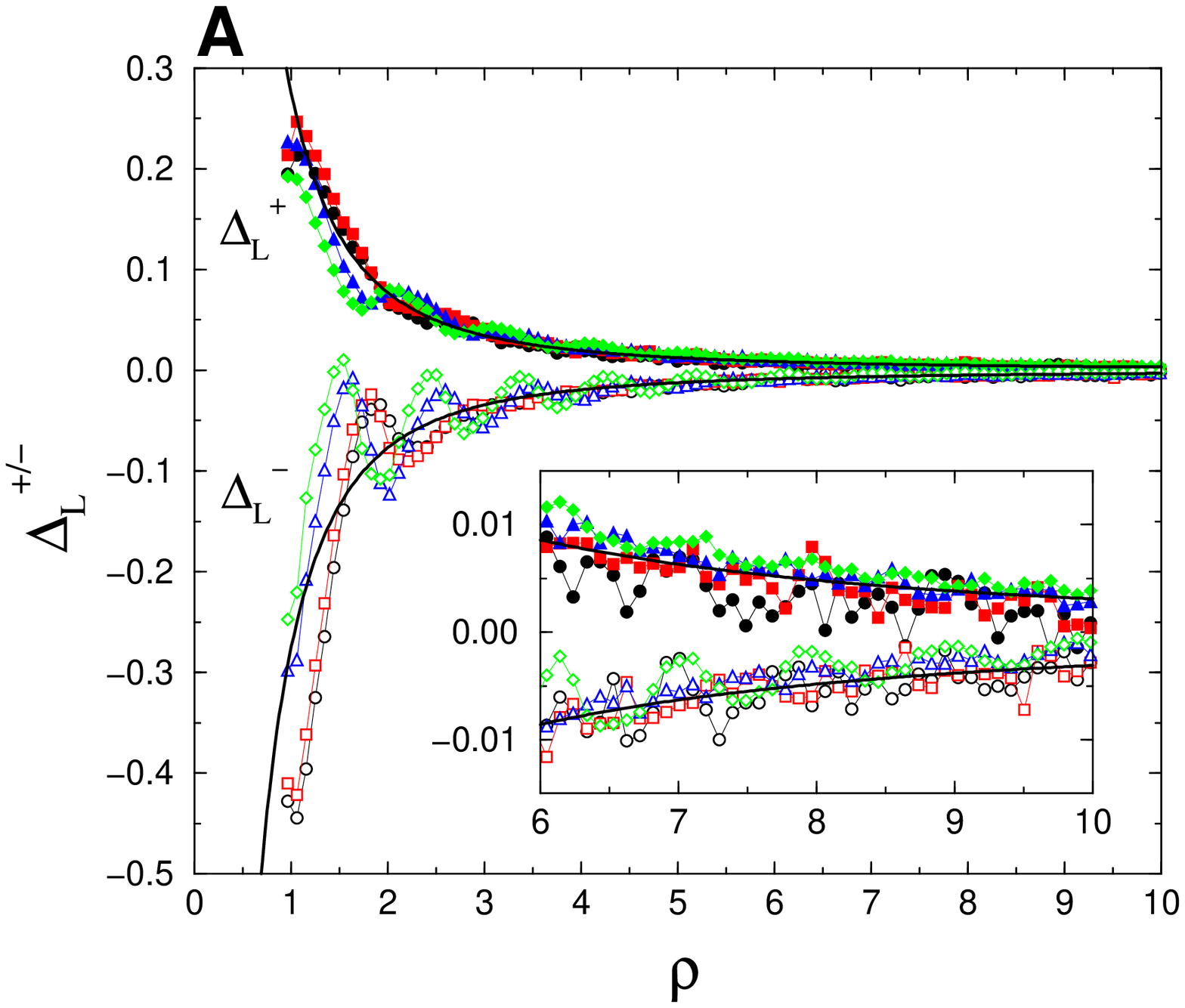}}}
\vspace{-0.5cm}
\centerline{
\resizebox{0.4\textwidth}{!}
{\includegraphics{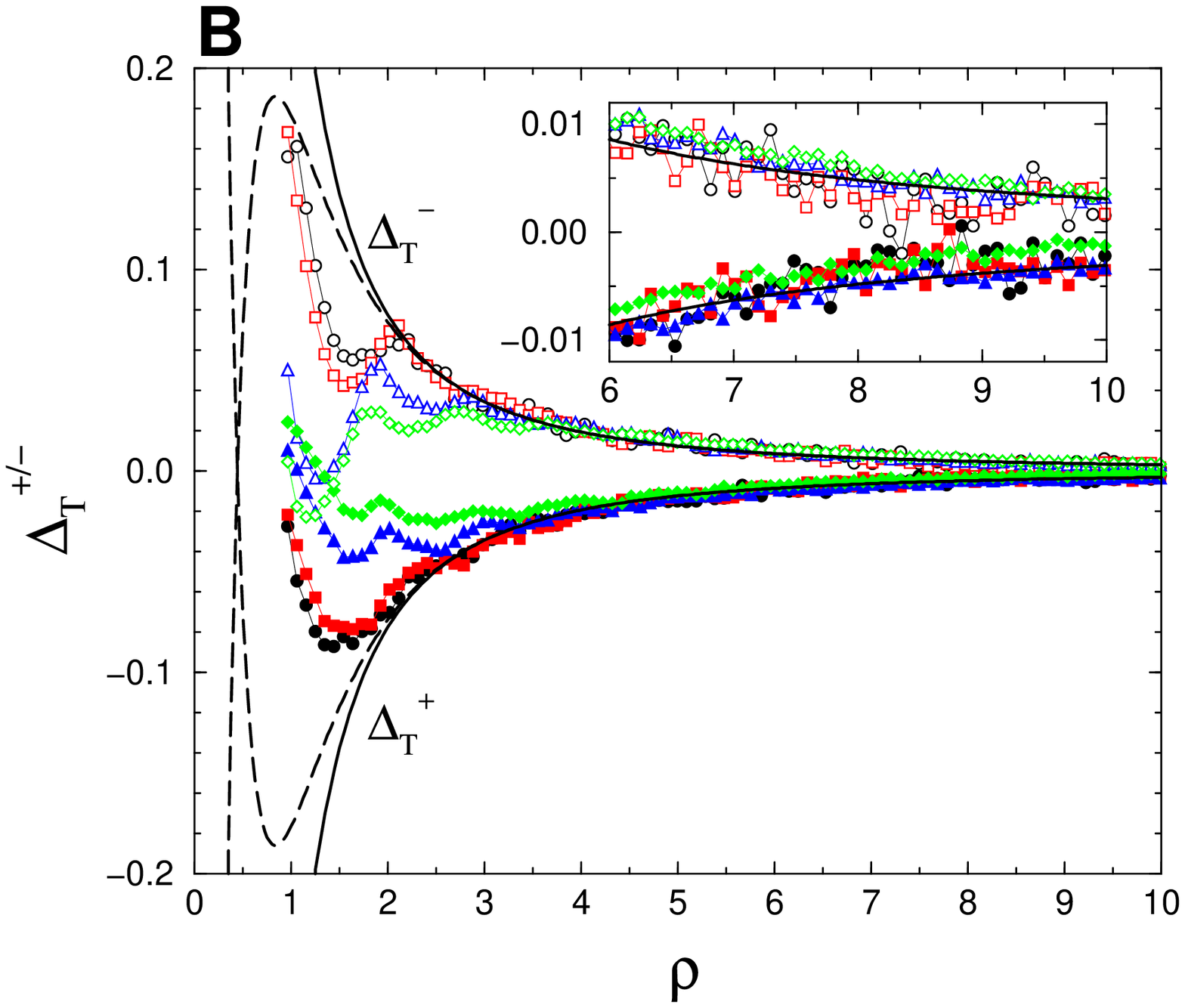}}}
\vspace{-.5cm}
\caption[]{(Color online)
The same data as in Fig.\ 2 rescaled according to Eq.\
(\ref{Delta}). Note the collapse of the long-distance data for
different area fractions onto the same four curves (insets). The solid
curves correspond to $\pm\lambda/\rho^2$ with $\lambda=0.31$ [Eq.\
(\ref{Delta_asym})].  The dashed curves in (b) are obtained from the
rigorous solution in the limit $a\ll w$ \cite{LM} (see text).}
\label{fig_Delta}
\end{figure}

Figure \ref{fig_D} shows the measured values of $D_{\rmL,\rmT}^\pm(r)$
as functions of $r$ at different area fractions $\phi$. As $\phi$ is
increased the curves shift to lower values (\ie $D_\rms$ expectedly
decreases with concentration) and, at short distances, become more
structured \cite{note_wiggles}. We focus in this Letter on the
long-distance pair interaction. To get a more unified presentation in
this domain, we rescale diffusion coefficients and distances according
to
\begin{equation}
  \Delta_{\rmL,\rmT}^\pm(\rho) \equiv 
  (D_{\rmL,\rmT}^\pm - D_\rms)/(D_0a/w),\ \ \rho\equiv r/w,
\label{Delta}
\end{equation}
where $D_0\equiv k_\rmB T/(6\pi\eta a)$ is the self-diffusivity of an
isolated, unconfined sphere ($k_\rmB T$ being the thermal energy and
$\eta$ the viscosity). This rescaling yields, in the limit of
small particles and low concentration, parameter-free functions 
describing solely the pair interaction
\cite{PRL02}.
The rescaled results are given in Fig.\ \ref{fig_Delta}, where all
long-distance measurements collapse onto the same four
curves.

In the longitudinal modes, the correlation has the expected positive
sign---particles drag one another in the direction of motion, thus
enhancing the collective motion and suppressing the relative motion
[Fig.\ \ref{fig_Delta}(a)].  The transverse interaction, however, has
a surprising reverse sign---the collective motion is suppressed while
the relative motion is enhanced [Fig.\ \ref{fig_Delta}(b)], \ie
particles exert ``anti-drag'' on one another.

Since the liquid velocity vanishes on the plates
(assuming no-slip boundary conditions), one might have expected
suppression of locally induced flows and a 
cutoff in the hydrodynamic interaction at distances much larger than
$w$. Such screening is found in q-1D suspensions
\cite{PRL02} and in the electrostatic analogue
of charges located between two conducting plates.
However, all four diffusion
coefficients are found to decay as $1/\rho^2$ for large $\rho$ (Fig.\
\ref{fig_Delta}). This decay is faster than the $1/\rho$ dependence in
unconfined suspensions but is still long-ranged---correlations at
distances ten times $w$ are measurable. Moreover, the observed decay
is {\em slower} than the $1/\rho^3$ dependence of the interaction
parallel to a single wall \cite{2par1wall}.

Other particles obstruct the flow induced by the monitored pair, 
thus affecting their measured interaction, as
observed at short distances (Fig.\ \ref{fig_Delta})
\cite{note_wiggles}.  One would have expected that at
large distances the combined contributions from numerous particles
would shift the pair
interaction by a factor proportional to $\phi$. This
correction should have been significant already at small $\phi$
because of the long-range interaction.
However, as can be seen in the insets of 
Fig.\ \ref{fig_Delta}, within our experimental accuracy there is {\it
no} concentration effect even at large $\phi$
\cite{note_Ds}.

Since the Reynolds number is very low ($\sim 10^{-6}$),
the hydrodynamic interactions are carried by linear Stokes flows
\cite{Happel}. Together with no-slip boundary conditions on the
plates and particles the mathematical problem is well posed, yet very
complicated to solve. The observations presented above can be
accounted for, nonetheless, on simpler grounds.
We focus on the flow due to the motion of a single particle parallel
to the plates. The motion perturbs the liquid momentum distribution
and displaces liquid mass. Far away from the particle one can consider
only the lowest-order moments of the perturbation---a momentum
monopole and a mass dipole. (Since mass is merely displaced, there is
no mass monopole.)  The momentum monopole undergoes diffusion with
absorbing boundary conditions at the plates, and its
contribution to the far field is therefore exponentially small
in $r/w$. However, the mass-dipole perturbation is not absorbed by the
boundaries and must propagate laterally. We thus expect the far field
to be that of a {\it two-dimensional mass dipole} (source doublet),
which decays only as $1/r^2$. The entrainment of a second particle by
this flow determines the pair hydrodynamic interaction, hence the
$1/r^2$ decay in the measured correlations.  Dipolar flow fields were
previously encountered in various q-2D and 2D problems
\cite{LM,Ajdari,flowsimul}.
The above argument clarifies the generality of
this result.

\begin{figure}[tbh]
\centerline{\resizebox{0.23\textwidth}{!}
{\includegraphics{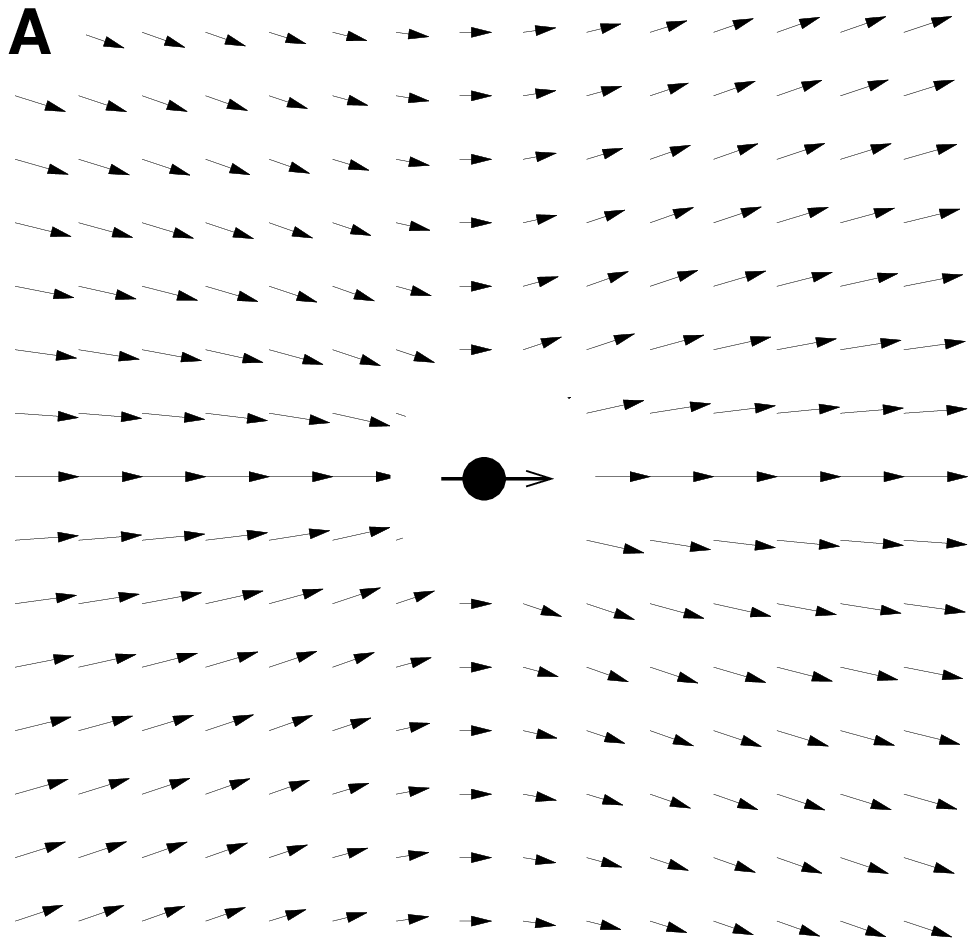}}
\hspace{0.01\textwidth}
\resizebox{0.23\textwidth}{!}
{\includegraphics{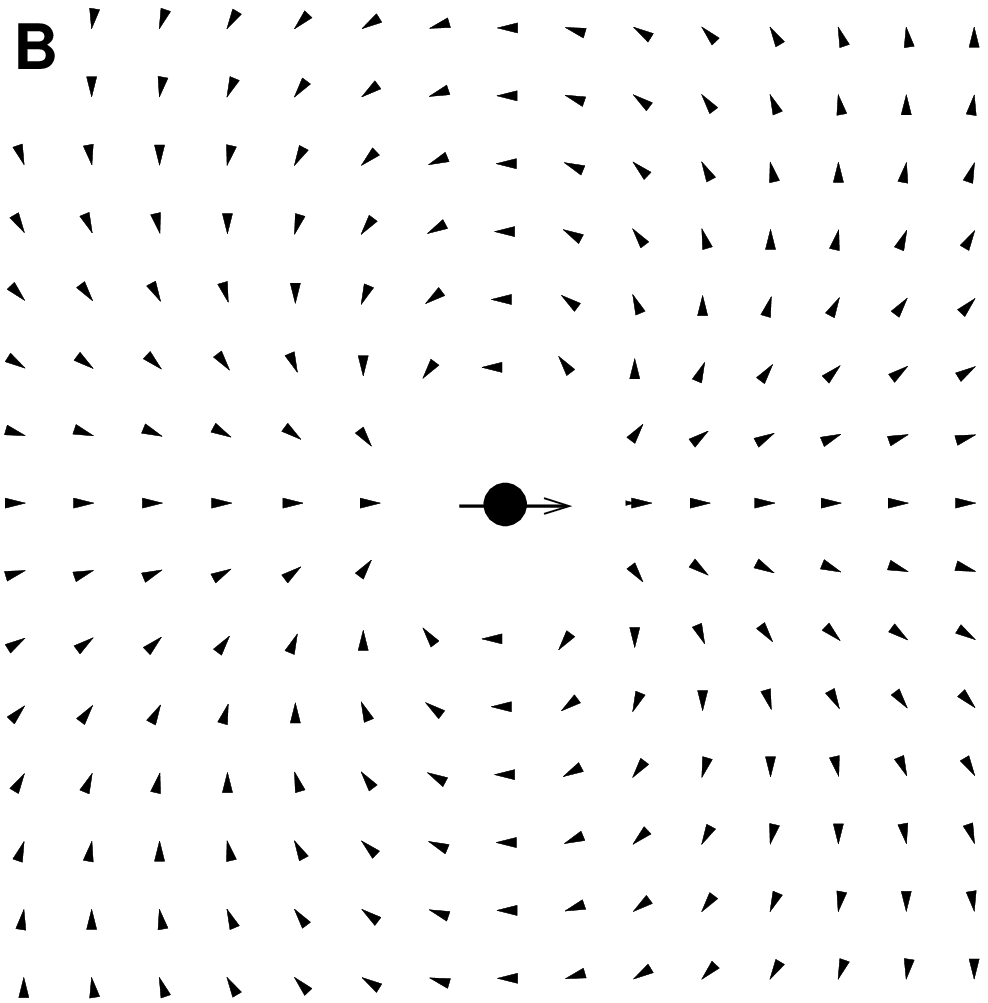}}}
\caption[]{Flow velocity field 
away from a particle moving to the right (indicated by circle and
arrow at the center), in (a) unconfined liquid \cite{Happel} and (b)
liquid confined between two plates [Eq.\ (\ref{flow})]. Shown are
cross-sections of the three-dimensional fields along the plane of
particle motion. Note the dipolar shape of the confined flow.}
\label{fig_flow}
\end{figure}

Consider a particle located on the midplane and exerting a force
$f_1\hat{\vecx}$ on the liquid.  The $y$ axis is taken as the second
direction parallel to the plates, and $z$ perpendicular to the plates
[Fig.\
\ref{fig_system}(b)]. The far velocity field, arising from a 
two-dimensional mass dipole oriented along $x$, 
is given by $v_\alpha(\vecr)=[f_1/(6\pi\eta w)]\Delta_\alpha(\vecr)$,
with
%
%
\begin{equation}
  \Delta_x(\vecr) = h(z) \frac{x^2-y^2}{(x^2+y^2)^2},\ 
  \Delta_y(\vecr) = h(z) \frac{2xy}{(x^2+y^2)^2},
\label{flow}
\end{equation}
while $\Delta_z$ is exponentially small in $r/w$. The flow has a
perpendicular profile $h(z)$ having a certain value at the midplane,
$h(0)\equiv w^2\lambda$, and vanishing on the two plates, $h(z=\pm
w/2)=0$. (In the limit $a/w\ll 1$ $h(z)$ is parabolic and
$\lambda=9/16$ \cite{LM}.)  The dipolar flow pattern is depicted in
Fig.\ \ref{fig_flow}(b), where it is contrasted with the flow in an
unconfined medium [Fig.\ \ref{fig_flow}(a)].  Due to the circulation
currents a second particle, located transverse to the particle
motion, will evidently be entrained
in the opposite direction, hence the negative transverse correlation.

To leading order in $a/r$, the longitudinal and transverse pair interactions
are given, respectively, by $\Delta_x(r\hat{\vecx})$ and 
$\Delta_x(r\hat{\vecy})$. 
Thus, from Eq.\ (\ref{flow}),
\begin{equation}
  \Delta_\rmL^\pm(\rho\gg 1) = \pm{\lambda}/{\rho^2},\ \ \  
  \Delta_\rmT^\pm(\rho\gg 1) = \mp{\lambda}/{\rho^2}.
\label{Delta_asym}
\end{equation}
This result is in good agreement with the data of Fig.\
\ref{fig_Delta} using $\lambda(a/w\simeq 0.45)\simeq 0.31$.  In Fig.\
\ref{fig_Delta}(b) we present also the results of the full solution in
the limit $a\ll w$ \cite{LM}, multiplied by $0.31/(9/16)=0.55$ to
match the long-distance behavior. This calculation
reproduces the nonmonotone behavior of $\Delta_{\rmT}^+$ at short
distances \cite{note_long}.

We now outline the analysis of the three-body effect \cite{ournext}.
Particle 1 exerts a force $f_1\hat{\vecx}$, yet now in
the presence of particle 2 at $\vecr$ and
particle 3 at $\vecr'$, where both $r,r'\gg w$ [Fig.\
\ref{fig_system}(b)]. 
The leading effect of particle 3 on the pair interaction between 1
and 2 arises from variations of the induced flow over the volume it
occupies. Because of symmetries of the midplane and Eq.\ (\ref{flow})
we find that the only derivative contributing to this effect is
$a^2\partial^2v_\alpha/\partial z^2 |_{z=0}\sim
(a^2/w^2)v_\alpha$ \cite{ft_Faxen}. We can therefore write the
force exerted by particle 3 as
$
  f_{3,\alpha} = (6\pi\eta a^3/w^2) C_{\alpha\beta} v_\beta(\vecr')
$,
where $C_{\alpha\beta}$ are dimensionless coefficients which, in
principle, could be obtained from a detailed calculation.  From
rotational symmetry $C_{xx}=C_{yy}\equiv C_\rmL$ and
$C_{xy}=-C_{yx}\equiv C_\rmT$. The force $\vecf_3$ perturbs the
velocity of particle 2 in the $x$ direction by
$
  \delta v_x(\vecr)=(6\pi\eta w)^{-1}\Delta_\alpha(\vecr-\vecr')
  f_{3,\alpha}
$,
and the perturbation should be averaged over all
third-particle positions $\vecr'$. The inter-particle distances being much
larger than $a$, we can assume a uniform probability per unit area, 
$\phi/(\pi a^2)$, to find a third particle at $\vecr'$. 
Combining this with the results for $\vecf_3$ and $\delta v_x$
and rescaling, we find the following corrections to the
pair interactions:
\begin{eqnarray}
  \delta\Delta_{\rmL,\rmT}^\pm(\rho) &=& \pm({a}/{\pi w})\phi C_\rmL 
  \int\rmd^2\rho'[\Delta_x(\vecrho-\vecrho')\Delta_x(\vecrho') 
  + \nonumber\\
  & & \Delta_y(\vecrho-\vecrho')\Delta_y(\vecrho')],
\label{conv}
\end{eqnarray}
where $\delta\Delta_\rmL^\pm$ and $\delta\Delta_\rmT^\pm$ are obtained
by setting, respectively, $\vecrho=\rho\hat{\vecx}$ and 
$\vecrho=\rho\hat{\vecy}$.
The integral in Eq.\ (\ref{conv}), however, vanishes. 
%
The leading corrections to this result that are linear in
$\phi$ are found to decay faster than $1/\rho^2$
\cite{ournext} and, hence, the vanishing of the three-body term is
asymptotically exact.

This approach can be extended to the four-body effect
\cite{ournext}, resulting in finite, small corrections.
We find, \eg
$
 \delta\Delta_\rmL^\pm/\Delta_\rmL^\pm = 
 (1/8\pi)\lambda^2(C_\rmL^2+C_\rmT^2)(a/w)^2\phi^2
$.
At our largest $\phi$ this amounts to $\sim(3\times
10^{-4})(C_\rmL^2+C_\rmT^2)$ which (unless $C_{\rmL,\rmT}$ are
unexpectedly large) would be smaller than the experimental noise.  For
most practical purposes the hydrodynamic interaction at large
distances is thus a long-ranged, yet purely pairwise effect even at
high concentrations. This result is unique to the q-2D geometry.

Perpendicular particle motion is restricted in our system by the large
$a/w$ ratio and possible repulsion from the plates \cite{JCP02}. The
flow induced by such motion decays exponentially with
$r/w$. Increasing the spacing $w$ is expected, therefore, to 
shift the validity range of our results to larger inter-particle
distances. This effect, however, requires further experimental study.
Since diffusion coefficients are
proportional to mobility coefficients, the results presented here for
Brownian motion should apply as well to overdamped driven motion, as
in sedimentation
\cite{sediment} or microfluidic transport
\cite{Whitesides}.

\begin{acknowledgments}
We thank Tom Witten for a helpful insight. We benefited from
discussions with Armand Ajdari, Leo Kadanoff, Sidney Nagel, Alexei
Tkachenko, and Wendy Zhang. This research was supported by the
National Science Foundation (CTS-021774 and CHE-9977841) and the
NSF-funded MRSEC at The University of Chicago.  H.D.\ acknowledges
support from the Israel Science Foundation (77/03) and the Israeli
Council of Higher Education (Alon Fellowship).
\end{acknowledgments}




\begin{thebibliography}{99}
\bibitem{note_w}
The value of $w$ is determined from the measured self-diffusivity of
single particles \cite{1par2wall}.
\bibitem{note_long}
For $\Delta_\rmL^\pm$ the results of this calculation are
indistinguishable (within the experimental range of distances) from
the asymptotic $\pm\lambda/\rho^2$ curves.
\bibitem{ft_Faxen}
This can be viewed as a modified Faxen's first law
\cite{Russel} for the two-plate case. In this geometry
one has $f_{3,\alpha}\sim [C'_{\alpha\beta}(\partial^2/\partial x^2 +
\partial^2/\partial y^2) + C''_{\alpha\beta}\partial^2/\partial z^2]
v_\beta$, with $C', C''$ two coefficient tensors.
Yet the field of Eq.\ (\ref{flow}) has a vanishing $xy$
Laplacian, and we are left with the last term.
\end{thebibliography}
\end{document}